\documentclass[aps,twocolumn,prd,showpacs,nofootinbib,showkeys,superscriptaddress,preprintnumbers]{revtex4-1}

\usepackage{graphicx,floatflt}
\usepackage{amsmath}
\usepackage{amstext}
\usepackage{epsfig}
\usepackage[usenames]{color}
\usepackage[dvipsnames]{xcolor}

\newcommand{\dd}{\mathrm{d}} 
\newcommand{\Mpl}{M_{\rm P}} 
\newcommand{\mpl}{m_{\rm P}} 

\newcommand{\epsone}{\epsilon_1}
\newcommand{\epstwo}{\epsilon_2}
\newcommand{\rr}{\mathrm}
\newcommand{\ns}{n_{\rr s} }

\newcommand{\phic}{\phi_{\rr c}}
\newcommand{\psic}{\psi_{\rr c}}

\newcommand{\Nend}{N_{\rr {end}}}

\newcommand{\xiend}{\xi_{\mathrm{end}}}

\newcommand{\fNL}{f_{\mathrm{NL}}}
\newcommand{\NN}{{\cal N}}

\newcommand{\Mpc}{\rr{Mpc}}

\newcommand{\msun}{M_\odot}

\newcommand{\mPBH}{M_{\rr{PBH}}}
\newcommand{\kg}{\rr{kg}}

\newcommand{\be}{\begin{equation}}
\newcommand{\ee}{\end{equation}}
\newcommand{\ba}{\begin{eqnarray}}
\newcommand{\ea}{\end{eqnarray}}

\begin{document}

\preprint{}

\title{Massive Primordial Black Holes from Hybrid Inflation \\ as Dark Matter and the seeds of Galaxies}

\author{S\'ebastien Clesse} 
\email{sebastien.clesse@unamur.be}
\affiliation{Namur Center of Complex Systems (naXys), Department of Mathematics, University of Namur, Rempart de la Vierge 8, 5000 Namur, Belgium}
\author{Juan Garc\'ia-Bellido} \email{juan.garciabellido@uam.es} 
\affiliation{Instituto de F\'isica Te\'orica UAM-CSIC, Universidad
Auton\'oma de Madrid, Cantoblanco, 28049 Madrid, Spain}

\date{\today}

\begin{abstract}
In this paper we present a new scenario where massive Primordial Black Holes (PBH) are produced from the collapse of large curvature perturbations generated during a mild waterfall phase of hybrid inflation.   We determine the values of the inflaton potential parameters leading to a PBH mass spectrum peaking on planetary-like masses at matter-radiation equality and producing abundances comparable to those of Dark Matter today, while the matter power spectrum on scales probed by CMB anisotropies agrees with {\sc Planck} data.  These PBH could have acquired large stellar masses today, via merging, and the model passes both the constraints from CMB distortions and micro-lensing.  This scenario is supported by {\sc Chandra} observations of numerous BH candidates in the central region of Andromeda.  Moreover, the tail of the PBH mass distribution could be responsible for the seeds of supermassive black holes at the center of galaxies, as well as for ultra-luminous X-rays sources.   We find that our effective hybrid potential can originate e.g. from D-term inflation with a Fayet-Iliopoulos term of the order of the Planck scale but sub-planckian values of the inflaton field.  Finally, we discuss the implications of quantum diffusion at the instability point of the potential, able to generate a swiss-cheese like structure of the Universe, eventually leading to apparent accelerated cosmic expansion. 
\end{abstract}
\pacs{98.80.Cq}
\maketitle

\section{Introduction}

A major challenge of present-day cosmology is the understanding of the nature of dark matter, accounting for about thirty percent of the total energy density of the Universe.  Among  a large variety of models, it has been proposed that dark matter is composed totally or partially by Primordial Black Holes (PBH) \cite{Carr:1974nx,Carr:1976zz,Carr:1975qj,Khlopov:2008qy,Frampton:2010sw,Blais:2002nd}.  These are formed in the early Universe when sufficiently large density fluctuations collapse gravitationally.  A threshold value of $\delta \rho /\rho \sim \mathcal O(1)$ is a typical requirement to ensure that gravity overcomes the pressure forces~\cite{Harada:2013epa,Musco:2012au,Musco:2008hv,Polnarev:2006aa,Musco:2004ak,Shibata:1999zs,Niemeyer:1999ak,Niemeyer:1997mt,Nakama:2013ica}. 

Several mechanisms can lead to the formation of PBH, e.g. sharp peaks in density contrast fluctuations generated during inflation~\cite{GarciaBellido:1996qt}, first-order phase transitions~\cite{Jedamzik:1999am}, resonant reheating~\cite{Suyama:2004mz}, tachyonic preheating~\cite{Suyama:2006sr} or some curvaton scenarios~\cite{Kohri:2012yw,Kawasaki:2012wr,Bugaev:2013vba}.   Large curvature perturbations on smaller scales than the ones probed by CMB anisotropy experiments can also be generated during inflation~\cite{Drees:2011yz,Drees:2012sz,Frampton:2010sw,Blais:2002nd,Erfani:2013iea,Drees:2011hb,Kawaguchi:2007fz,Kohri:2007qn}, e.g. for hybrid models ending with a fast (in terms of $e$-folds of expansion) waterfall phase.   In this case, exponentially growing modes of a tachyonic auxiliary field induce order one curvature perturbations~\cite{GarciaBellido:1996qt,Lyth:2011kj,Bugaev:2011wy} and PBH can be formed when they re-enter inside the horizon during the radiation era.  However, in the standard picture of hybrid inflation, the corresponding scales re-enter into the horizon shortly after the end of inflation, leading to the formation of PBH with relatively low masses, $\mPBH \lesssim  \mathcal O(10)~\kg$.  These PBH evaporate in a very short time, compared to the age of the Universe, and cannot contribute to dark matter today.   This process can nevertheless eventually contribute to the reheating of the Universe~\cite{GarciaBellido:1996qt}.

Tight constraints have been established on PBH mass and abundance from various theoretical arguments and observations, like the evaporation through Hawking radiation, gamma-ray emission, abundance of neutron stars, microlensing and CMB distortions.  It results that PBH cannot contribute for more than about 1\% of dark matter, except in the range $10^{18}~\kg \lesssim \mPBH \lesssim 10^{23}~\kg$, as well as for masses larger than around a solar mass, $M\gtrsim \msun \sim10^{30}~\kg$, under the condition that they do not generate too large CMB distortions.  It is also unclear whether some models predicting a broad mass spectrum of PBH can be accommodated with current constraints, while generating the right amount of dark matter when integrated over all masses.  

In this paper, we present a new scenario in which the majority of dark matter consists of PBH with a relatively broad mass spectrum covering a few order of magnitudes, possibly up to $\mathcal O(100)$ solar masses.  The large curvature perturbations at the origin of their formation are generated in the context of hybrid inflation ending with a mild waterfall phase.  This is a regime where inflation continues for several $e$-folds (up to 50 $e$-folds) of expansion during the final waterfall phase.  Compared to the standard picture of fast waterfall, important curvature perturbations are generated on larger scales, that reenter into the horizon at later times and thus lead to the formation of PBH with larger masses.

More precisely, we consider for inflation an effective hybrid-like two-field potential, having a nearly flat valley where field trajectories are slowly evolving when scales relevant for CMB anisotropies exit the Hubble radius.  In order to avoid a blue spectrum of scalar perturbations, which is a generic prediction of the original hybrid model in the false vacuum regime~\cite{Linde:1993cn,Copeland:1994vg}, the effective potential has a negative curvature close to the critical point below which the tachyonic instability develops~\cite{Felder:2000hj}.  The slope of the potential at this critical point is nevertheless sufficiently small for the final waterfall phase to be mild and to last for typically between $10$ and $50$ $e$-folds of expansion. 
In this scenario, the second slow-roll parameter gives the dominant contribution to the scalar spectral index, which can accommodate the recent constraints from Planck~\cite{Ade:2013lta,Ade:2013rta}.   This scenario is similar to the case of a mild waterfall phase with more than 50 $e$-folds of expansion during the waterfall~\cite{Clesse:2010iz,Clesse:2013jra,Kodama:2011vs,Clesse:2012dw,Mulryne:2011ni}, but it must be seen as a transitory case because observable scales exit the horizon prior to the waterfall.   The present scenario has therefore strong similarities with the ones studied in Refs~\cite{Abolhasani:2010kn,GarciaBellido:1996qt}.  

In the context of a long waterfall phase, during the first stage of the waterfall when the field dynamics is still governed by the slope of the potential in the direction of the valley, entropy perturbations due to the presence of the auxiliary field grow exponentially and source the power spectrum of curvature perturbations whose amplitude can grow up to values larger than unity~\cite{Clesse:2013jra}.  Then in the second phase of the waterfall, field trajectories reach an attractor and are effectively single field, so that curvature perturbations becoming super-Hubble at this time fall back to much lower values. Depending on the model parameters, one predicts a broad peak in the power spectrum of curvature perturbations, whose maximal amplitude can exceed the threshold value for the formation of PBH.  

By numerically solving the multi-field homogeneous and linear perturbation dynamics during the waterfall, and by cross-checking our result using the $\delta N$ formalism (both analytically and numerically), we calculate the power spectrum of curvature perturbations at the end of inflation.  Particular care is given to take into account the effect of quantum stochastic fluctuations of the auxiliary field at the instability point, which can significantly change the classical evolution during the waterfall.  From the spectrum of curvature perturbations, the formation process of PBH is studied and their mass spectrum is evaluated.  Finally, the contribution of PBH to the density of the Universe at matter-radiation equality is calculated, and we determine the parameter values of the inflationary potential leading to the right amount of dark matter.    As mentioned later, we find that those parameters can fit simultaneously with the observational constraints on the curvature power spectrum from CMB experiments.   We argue that micro-lensing and CMB distortion constraints can be naturally evaded if PBH are initially sub-solar and grow by merging after recombination until they acquire a stellar mass today.  As a result, the mass distribution of PBH could explain the excess of few solar mass black holes in Andromeda that has been observed recently~\cite{Barnard:2014afa,Barnard:2013nqa,Barnard:2013dea,Barnard:2012tn,Barnard:2011pv}.  Finally we discuss whether PBH in the tail of the mass spectrum can serve as the seeds of the supermassive black holes observed in galaxies and quasars at high redshifts~\cite{Fan:2003wd,Willott:2003xf,2010Natur.467..940L,Iye:2006mb,Oesch:2013pt,2014ApJ...786..108O}.  
 
The paper is organized as follows: In Sec.~\ref{sec:PBHconstraints} we review the principal constrains on PBH abundances.  The effective model of hybrid inflation with a mild waterfall is introduced in Sec.~\ref{sec:model}.  The field dynamics is described and the curvature power spectrum is calculated in this section.  The formation of PBH from large curvature perturbations produced during the waterfall is studied in Sec.~\ref{sec:PBHformation}, where we also derive their mass spectrum and their contribution to the density of the Universe at matter-radiation equality.  In Sec.~\ref{sec:infconstraints}, we identify the model parameter ranges leading to the right amount of dark matter, with mass spectra evading the observational constraints.  In Sec.~\ref{sec:embedding}, we discuss how to embed our effective inflationary potential in a realistic high-energy framework.  The level of CMB distortions induced by an excess of power on small scales is evaluated in Sec.~\ref{sec:distortions} and compared to the limits reachable by future experiments like PIXIE and PRISM.   In Sec.~\ref{sec:seeds} we discuss how the most massive PBH can be identified to the seeds of the supermassive black holes in quasars at high redshifts, and in the center of galaxies today.  We conclude and present some interesting perspectives in Sec.~\ref{sec:conclusion}.
 
\section{Primordial black holes and observational constraints}  \label{sec:PBHconstraints}

Because primordial black holes are non-relativistic and effectively nearly collisionless, they are good candidates for dark matter.  The mass spectrum of primordial black holes is nevertheless severely constrained by several types of observations, which are listed and briefly explained below (for a review, see Ref.~\cite{Carr:2009jm}).
\begin{enumerate}
\item \textit{Lifetime of primordial black holes:}  Due to the Hawking radiation, PBH evaporate on a time scale of the order of $t_{\rr{ev}}(M) \sim G^2 M^3 / \hbar c^4  $, so that PBH with a mass $\mPBH \simeq 5 \times 10^{11}$ kg evaporate in a time much shorter than the age of the Universe~\cite{Carr:1976zz,Carr:2009jm} and therefore cannot contribute significantly to dark matter today.   

\item \textit{Light element abundances:}  PBH evaporation may also have an effect on Big Bang Nucleosynthesis (BBN).  Only PBH of masses $\mPBH \lesssim 10^7~\kg $ evaporate before the BBN.   More massive ones can affect light element abundances through the emission of mesons and anti-nucleons; and through the hadro-dissociation and photo-dissociation processes.   Strong bounds have been established on PBH abundance from BBN~\cite{Carr:2009jm}, but those only concern PBH of masses $\mPBH \lesssim 10^{11}~\kg$ which are not good candidates for dark matter due to early evaporation.    

\item \textit{Extragalactic photon background:}  Evaporating PBH at the present epoch can emit observable extragalactic gamma-ray radiation.  The  photon intensity spectrum can be calculated.  For instance, the Hawking radiation produced by PBH with mass $\mPBH \simeq 10^{13}~\kg$ is responsible for the emission of $\sim 100$ MeV radiation, which should have been observed with the EGRET and Fermi Large Area telescopes if $\Omega_{\rr{PBH}} > 0.01$~\cite{Carr:2009jm}.   
PBH cannot account for the totality of dark matter if $\mPBH \lesssim 7 \times 10^{12}~\kg$.  Note however that this limit assumes a uniform distribution of PBH throughout the Universe, which is not realistic since dark matter clusters like galaxies, galaxy clusters and super-clusters.

\item \textit{Galactic background radiation:}   PBH are expected to have clustered with galactic halos, and thus there should be also a galactic background of gamma radiation, which should be anisotropic and in principle separable from the extra-galactic emission.   The constraints from galactic background radiation are close but less competitive than BBN and extra-galactic ones.  A distinctive signature of PBH could also be seen in the ratio of antiprotons to protons, in the energy range between 100 MeV and 10 GeV, in the galactic cosmic ray spectrum~\cite{Carr:2009jm}.  This gives typically similar constraints on the abundance of PBH.

\item \textit{Femtolensing of gamma-ray bursts:}  Compact objects can induce gravitational femto-lensing of gamma-ray bursts.  The lack of femto-lensing detection in the Fermi Gamma-Ray Burst Monitor experiment has provided evidence that in the mass range $5 \times 10^{14} - 10^{17}~\kg$, PBH cannot account for a large fraction of 
dark matter~\cite{Barnacka:2012bm}.  

\item \textit{Star Formation:}  If star formation occurs in an environment dominated by dark matter, constituted partially or totally of PBH, these can be captured by stars, they sink to the center, and at the end of the star evolution they destroy in a very short time by accretion the compact remnant (a white dwarf or a neutron star).   The constraints resulting from the observation of neutron stars and white dwarves in globular clusters does not allow PBH to constitute totally the dark matter in the mass range $10^{13}~\kg \lesssim \mPBH \lesssim 10^{19}~\kg$~\cite{Capela:2012jz}.   

\item \textit{Capture of PBH by neutron stars:}   In a similar way, PBH can be captured by neutron stars which are then accreted onto the PBH and destroyed in a short time.  Assuming large dark matter densities and low velocities, conditions that can be fulfilled in the cores of the globular clusters, PBH cannot account entirely for dark matter in the range $10^{15}~\kg < \mPBH < 10^{21}~\kg$~\cite{Capela:2013yf}.  However, the dark matter density inside globular clusters is not known, and those constraints (as well as constraints from star formation) are evaded if the dark matter density is $\rho_{\rr{DM}}^{\rr{Glob. Cl.}} \lesssim 10^2 \ \rr{GeV cm^{-3}}$.

\item \textit{Microlensing surveys:}   If the dark matter galactic halo is mostly composed of PBH, one expects gravitational micro-lensing events of stars in the Magellanic clouds.  The EROS micro-lensing survey and the MACHO collaboration did not observe such events and have put a limit on PBH abundance in the range $10^{23}~\kg < \mPBH < 10^{31}~\kg$ ~\cite{Tisserand:2006zx,Alcock:1998fx}.   The Kepler dada have permitted to extend this range down to $4 \times 10^{21}~\kg \lesssim \mPBH \lesssim 2 \times 10^{23}~\kg$~\cite{Griest:2013aaa}.

\item \textit{CMB spectral distortions:}  X-rays emitted by gas accretion near PBH modify the recombination history, which generates CMB spectral distortions and CMB temperature anisotropies~\cite{Ricotti:2007au}.  Distortions are strongly constrained by the COBE/FIRAS experiment.  It results that PBH of  $\mPBH > 10 \msun$ cannot contribute to more than a few percent of the dark matter; whereas solar mass PBH cannot contribute to more than about 10\% of dark matter.  However these bounds assume that PBH masses do not change with time.  But merging and accretion since recombination can in principle lead to PBH with masses significantly larger than $10 \msun$ today while being sub-solar before recombination, thus evading both micro-lensing and CMB constraints.

Another source of distortions comes from super radiant instabilities of non-evaporating PBH, due to the release with the expansion of the energy associated to exponentially growing photon density around PBH.   The non-detection of distortions puts limits in the range $10^{22}~\kg \lesssim  \mPBH \lesssim 10^{29}~\kg$  for dark matter consisted of maximally-spinning PBH~\cite{Pani:2013hpa}.  

\end{enumerate}

\begin{figure}[!ht]
\begin{center} \includegraphics[height=60mm]{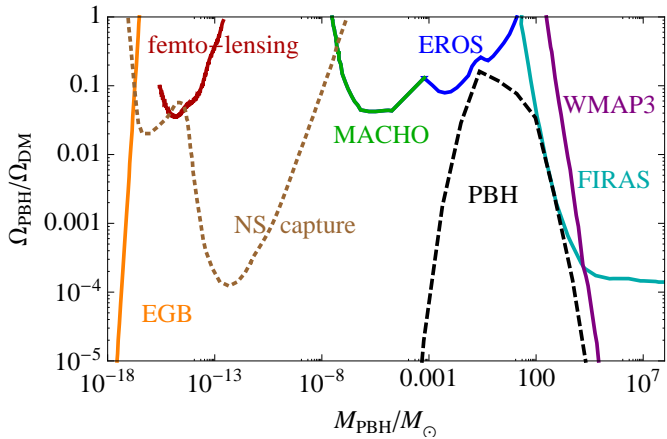}
\caption{\label{fig:PBHconstraints}
Limits on the abundance of PBH today, from extragalactic photon background (orange), femto-lensing (red), micro-lensing by MACHO (green) and EROS (blue) and CMB distortions by FIRAS (cyan) and WMAP3 (purple).  The constraints from star formation and capture by neutron stars in globular clusters are displayed for $\rho_{\rr{DM}}^{\rr{Glob. Cl.}} = 2 \times 10^3 \ \rr{GeV cm^{-3}}$ (brown). The  black dashed line corresponds to a particular realization of our scenario of PBH formation.  Figure adapted from~\cite{Capela:2013yf}. 
}
\end{center}
\end{figure}

The principal constraints on the fraction of dark matter due to massive PBH are summarized on Fig.~\ref{fig:PBHconstraints}.  The bounds from star formation and capture by neutron stars are displayed as brown dotted lines, since they are easily evaded if the dark matter density inside globular clusters is sufficiently low.   In this case, there exists a gap between $10^{17}~\kg \lesssim \mPBH \lesssim 10^{23}~\kg$ where there is no constraint, and thus where PBH can be identified to the dark matter component.   Finally, note that most limits have been obtained under the implicit assumption of a mono-chromatic distribution of PBH.  In the scenario proposed in this paper, the PBH mass spectrum is very broad, covering typically five orders of magnitudes, and it is still rather unclear how the current constraints must be adapted for such broad mass spectra.  

In the future, those constraints should be improved by other experiments and observations:  PBH passing through stars can lead to detectable seismic signatures (in the form of solar oscillations), induced by the squeeze of the star in presence of the PBH gravitational field.  PBH of masses larger than $10^{18}~\kg$ are potentially observable~\cite{Kesden:2011ij}.   Even if highly unlikely (1 event in $\sim 10^7 $ years for $\rho_{\rr{PBH}} = \rho_{\rr{DM}}$ with $\mPBH \sim 10^{12}~\kg$), the transit of PBH of masses $\mPBH \gtrsim 10^{12}~\kg$ through or nearby the Earth could be detected because of the seismic waves they induce~\cite{Luo:2012pp}.    X-rays photons emitted by non-evaporating PBH should ionize and heat the nearby intergalactic medium at high redshifts.   This produces specific signatures in the 21cm angular power spectrum from reionization, which could be detected with the SKA radio-telescope~\cite{Tashiro:2012qe}.   For PBH of masses from $10^2 \msun$ to $10^8 \msun$, densities down to $\Omega_{\rr{PBH}}\gtrsim 10^{-9}$ could be seen.   A PBH transiting nearby a pulsar gives an impulse acceleration which results in residuals on normally orderly pulsar timing data~\cite{Seto:2007kj,Kashiyama:2012qz}.  Those timing residuals could be detected with future giant radio-telescope like the SKA.  The signal induced by PBH in the mass range $10^{19}~\kg \lesssim \mPBH \lesssim 10^{25}~\kg$ and contributing to more than one percent to dark matter should be detected~\cite{Kashiyama:2012qz}.   Binaries of PBH forming a fraction of DM should emit gravitational waves; this results in a background of gravitational waves that could be observed by LIGO, DECIGO and LISA~\cite{2009arXiv0909.1738H,Dolgov:2013pha}.

Finally, the recent discovery by CHANDRA of tens of black hole candidates in the central region of the Andromeda (M31) galaxy~\cite{Barnard:2014afa,Barnard:2013nqa,Barnard:2013dea,Barnard:2012tn,Barnard:2011pv} provides a hint in favor of  models of PBH with stellar masses.  As detailed later in the paper, such massive PBH can be produced in our model.  The CMB distortions and micro-lensing limits could be evaded if PBH were less massive at the epoch of recombination and then have grown mostly by merging to form black holes with stellar masses today.   
  
\section{Hybrid-Waterfall inflation}  \label{sec:model}

It has been shown recently that the original non-supersymmetric hybrid model~\cite{Linde:1993cn,Copeland:1994vg} and its most well-known supersymmetric realizations, the F-term and D-term models~\cite{Dvali:1994ms,Binetruy:1996xj}, own a regime of mild waterfall~\cite{Clesse:2010iz,Kodama:2011vs,Clesse:2012dw,Clesse:2013jra,Mulryne:2011ni}.  Initially the field trajectories are slowly rolling along a nearly flat valley of the multi-field potential.  When trajectories cross a critical field value, denoted $\phic$, the potential becomes tachyonic in the orthogonal direction to the valley.  In the mild-waterfall case, inflation continues for more than 50 $e$-folds of expansion after crossing the critical instability point and before tachyonic preheating~\cite{Felder:2000hj} is triggered.    This scenario has the advantage that topological defects formed at the instability point are stretched outside our observable patch of the Universe by the subsequent inflation. 

According to Refs.~\cite{Kodama:2011vs,Clesse:2012dw,Clesse:2013jra}, the mild waterfall can be decomposed in two phases (called phase-1 and phase-2).  During the first one, inflation is driven only by the inflaton, whereas the terms involving the auxiliary field can be neglected.  At some point, these terms become dominant and trajectories enter in a second phase.
When the waterfall lasts for much more than 50 $e$-folds, the observable scales exit the Hubble radius in the second phase, when trajectories are effectively single field and curvature perturbations are generated by adiabatic modes only.   For the three hybrid models mentioned above (original, F-term and D-term), the observable predictions are consequently modified and a red scalar spectral index is predicted (instead of a blue one for the original model followed by a nearly instantaneous waterfall).  If one denotes by $N_*$ the number of $e$-folds between horizon exit of the pivot scale $k_* = 0.05 †\Mpc^{-1}$ and the end of inflation, the scalar spectral index is given by $\ns = 1 - 4/ N_*$, too low for being within the 95\% C.L. limits of Planck.  Only a low, non-detectable, level of local non-gaussianitiy is produced, characterized by $\fNL \simeq -1/N_*$~\cite{Clesse:2013jra}.  

When inflation continues during the waterfall for a number of $e$-folds close but larger than 50 $e$-folds, the pivot scale becomes super-Hubble during the phase-1.  Trajectories are not effectively single-field, and entropic perturbations source the curvature perturbations~\cite{Clesse:2013jra}.  This leads to a strong enhancement in the scalar power spectrum amplitude, whose thus cannot be in agreement with observations.   

In this paper, we focus on an intermediate case, between fast and mild waterfall.  We consider the regime where inflation continues for a number of $e$-folds between about 20 and 40 after crossing the instability point.   There is a major difference with the previous case: observable scales leave the Hubble radius when field trajectories are still evolving along the valley, when the usual single-field slow-roll formalism can be used to derive accurately the observable predictions.  Nevertheless, in order to study the waterfall phase, we have also integrated numerically the full multi-field dynamics, both a the background and linear perturbation level.  

In the following, we introduce the field potential and derive the scalar power spectrum amplitude and spectral index on scales that are relevant for CMB anisotropies.  Then we study the waterfall phase and calculate the power spectrum of curvature perturbations on small scales, and we show that for some parameters, the enhancement of power is so important that it leads later to the formation of massive PBH.  

\subsection{Along the valley}

Since the original hybrid potential with a constant plus a quadratic term in $\phi$ predicts a blue-tilted scalar spectrum, we have considered a more general form for the effective potential close to the critical point of instability.  Contrary to the original hybrid model, it exhibits a negative curvature in order to generate a red spectrum, plus a linear term in $\phi$ to control the duration of the waterfall phase.  The embedding of this model in some high-energy frameworks will be discussed in Sec.~\ref{sec:embedding}.  The considered two-field potential reads 
\begin{eqnarray} \label{eq:potential}
V(\phi,\psi) & = &\Lambda \left[ \left(1-\frac{\psi^2}{M^2}\right)^2 + \frac{(\phi-\phic)}{\mu_1} \right. \nonumber \\ 
  & - & \left.  \frac{(\phi-\phic)^2}{\mu_2^2}+ \frac{2 \phi^2 \psi^2}{ M^2 \phic^2} \right]~.
\end{eqnarray}
Initially, inflation takes place along the valley $\psi = 0$.   As shown in~\cite{Clesse:2008pf,Clesse:2009ur,Clesse:2009zd}, there is no fine-tuning of initial fields hidden behind this assumption.  Below the critical value $\phic$, this potential develops a tachyonic instability, forcing the field trajectories to reach one of the global minima, located at $\phi = 0$, $\psi = \pm M$.   Apart from the negative curvature and the additional linear term, the potential is identical to the one of the original hybrid model.  

The slope and the curvature of the potential at the critical point are thus controlled respectively by the mass parameters $\mu_1$ and $\mu_2$.  We assume that $\mu_1$ is sufficiently large compared to $\mu_2$ for the slope along the valley to be constant over the range of scales going from scales relevant to CMB anisotropies down do scales that exit the Hubble radius at the critical instability point.   
At $\phi = \phic$, the slow-roll approximation is valid and the first and second Hubble-flow slow-roll parameters are given by
\be
{\epsone}_{\phic} = \frac{\Mpl^2}{2} \left( \frac{V'}{V} \right)^2 = \frac{\Mpl^2}{2 \mu_1^2},
\ee
\be
{\epstwo}_{\phic} = 2 \Mpl^2  \left[ \left( \frac{V'}{V}  \right)^2 - \frac{V"}{V}  \right]  = 2 \Mpl^2 \left( \frac{1}{\mu_1^2} + \frac{2}{\mu_2^2}   \right).
\ee
where $\Mpl$ is the reduced Planck mass and a prime denotes the derivative with respect to the field $\phi$.   In the regime of interest, $\mu_1 \gg \mu_2$ and the scalar spectral index, given by
\be
\ns = 1- 2 {\epsone}_* - {\epstwo}_*,
\ee
is dominated by the contribution of the second slow-roll parameter.   The star index denotes quantities evaluated at the time $t_* $ when $k_* = a(t_*) H(t_*)$ with $k_*= 0.05 †\Mpc^{-1}$ being the pivot scale used by Planck.  One thus gets
\be
\ns \simeq 1 - \frac{4 \Mpl^2}{\mu_2^2}~.
\ee
If the scalar spectral index is given by the best fit value from Planck~\cite{Ade:2013lta}, $\ns \simeq 0.9603$, one obtains  
\be
\mu_2 = \frac{2 \Mpl}{\sqrt{1-\ns}} \simeq 10 \Mpl~.
\ee
The scalar power spectrum amplitude is also measured by Planck and is given at the pivot scale by
\begin{eqnarray}
\mathcal P_\zeta(k_*) & = &  \frac{H_*^2}{8 \pi^2 \Mpl^2 {\epsone}_*} \\
& \simeq & \frac{\Lambda \mu_1^2}{12 \pi^2 \Mpl^6} \left( \frac{k_*}{k_{\phic}}  \right)^{\ns - 1}  \\
&= & 2.21 \times 10^{-9}.
\end{eqnarray}
The second equality is derived by using the Friedmann equation in slow-roll $H^2 \simeq V /3 \Mpl^2$.  This leads to a relation between the $\Lambda$ and $\mu_1$ parameters,
\be \label{eq:Lambda}
\Lambda = 2.21 \times 10^{-9} \times \frac{12 \pi^2 \Mpl^6}{\mu_1 ^2} \left( \frac {k_{\phic}} {k_*} \right)^{\ns - 1}.
\ee
Practically, since the duration of inflation depends on the waterfall dynamics, $k_* / k_{\phic} $ cannot be known before a first integration of the background dynamics.  One needs also to assume a reheating history.  We consider for simplicity the case of instantaneous reheating~\cite{GarciaBellido:1997wm}.  For the numerical implementation of $\Lambda$ we use the following procedure: (i) we first guess its value assuming $k_* = k_{\phic} $, (ii) we solve numerically the 2-field dynamics and find the corresponding $k_* / k_{\phic} $, (iii) Eq.~(\ref{eq:Lambda}) is used to guess a better value of $\Lambda$.  We proceed iteratively until the scalar power spectrum amplitude corresponds to the Planck measurement.  Usually three iterations give a good agreement, because the waterfall dynamics in influenced only slightly by $\Lambda$, as discussed thereafter.

\subsection{Waterfall phase}

\subsubsection{Quantum diffusion at the critical point}

Once the critical instability point has been crossed, the waterfall phase takes place.  We have assumed that the classical 2-field dynamics is valid from the critical point, given a tiny initial displacement from $\psi = 0$.  In a realistic scenario the quantum diffusion close to $\phic$ plays the role of displacing the auxiliary field.  This process can be more or less efficient in different patches of the Universe, having an effect on the subsequent classical dynamics during the waterfall and on the resulting scalar power spectrum.   It is therefore important to take properly into account the quantum diffusion of $\psi$.

The auxiliary field distribution over a spatial region is a Gaussian whose width at the critical point of instability can be calculated by integrating the quantum stochastic dynamics of $\psi$ \cite{GarciaBellido:1996qt,Clesse:2010iz,Clesse:2013jra}.  It reads
\be \label{eq:psi0}
\psi_0 \equiv \sqrt{\langle \psi^2 \rangle} = \left( \frac{\Lambda  \sqrt{2 \phic \mu_1} M}{96 \pi^{3/2} }   \right)^{1/2} ,
\ee
where the brackets denote averaging in real space.  It must be also noticed that quantum diffusion only plays a role very close to the critical instability point and that the classical dynamics is quickly recovered.  Quantum effects taking place after crossing $\phic $ actually influence only marginally the waterfall dynamics, and for simplicity these have been neglected.  But we have taken into account that different classical evolutions can emerge from various values of $\psic$.  

To do so, for each parameter set we consider, the classical two-field dynamics is integrated numerically over 200 realizations of $\psic$, distributed according to a Gaussian of width $\psi_0$.  Then the mean scalar power spectrum is obtained by averaging over all realizations.  Since each realization can be more or less efficient in producing PBH (which is a non-linear process), the same averaging procedure is applied for the calculation of the fraction of the Universe that collapses into PBH.   In order to illustrate the importance of this effect, the scalar power spectrum for each of these realizations has been plotted on Fig.~\ref{fig:Pzeta} in addition to the averaged spectrum and the one obtained assuming $\psic = \psi_0$.

Finally, note that the inflaton field $\phi$ remains classical during the waterfall phase and drives the Universe expansion in the regime where the stochastic dynamics of $\psi$ can be important.   For a more precise investigation of the stochastic dynamics in hybrid inflation, the interested reader can refer to Ref.~\cite{Martin:2011ib}.

\subsubsection{Background classical dynamics}

In order to get the classical field trajectories and the expansion during the waterfall, one needs to solve the two-field dynamics governed by the Friedmann-Lema\^itre equation
\be
H^2 = \frac{1}{3 \Mpl^2} \left[ \frac{\dot \phi^2}{2} + \frac{\dot \psi^2}{2} + V(\phi,\psi)  \right]~,
\ee
and the Klein-Gordon equations
\be
\ddot \phi + 3 H \dot \phi + \frac{\partial V}{ \partial \phi} = 0,
\ee
\be
\ddot \psi + 3 H \dot \psi + \frac{\partial V}{ \partial \psi} = 0.
\ee
These exact equations have been implemented numerically.  But it is also possible to derive accurate analytical solutions, by considering the usual slow-roll approximation where the kinetic terms and the second field derivatives can be neglected.    

As already mentioned, the waterfall can be decomposed in two phases, called phase-1 and phase-2.  The slow roll dynamics of both fields can be integrated in each of them.  Let us first introduce the notation 
\be
\phi \equiv \phic \rr e^\xi, \hspace{1cm} \psi \equiv \psi_0 \rr e^\chi~.
\ee
During the waterfall, as long as the slow-roll approximation is valid, $| \xi| \ll 1$.   One therefore has $\phi \simeq \phic (1 + \xi)$ and the Klein Gordon equations for the scalar fields in the slow-roll approximation reduce to
\be  \label{eq:KG1}
3 H \dot \xi \simeq - \frac{2 \Lambda}{\mu_1^2} \left(1+\frac{2 \mu_1^2 \psi^2}{M^2 \phic^2}  \right),
\ee
\be \label{eq:KG2}
3 H \dot \chi  \simeq - \frac{4 \Lambda}{M^2} \left( 2 \xi + \frac{\psi^2}{M^2}  \right),
\ee
whereas the Friedmann-Lema\^itre equation is given by
\be
H^2 = \frac{\Lambda}{3 \Mpl^2}.
\ee
During phase-1, the second term of Eqs~(\ref{eq:KG1}) and~(\ref{eq:KG2}) are by definition negligible.  At some time, the second term in the r.h.s. of Eq.~(\ref{eq:KG1}) becomes larger than unity and the dynamics enters into phase-2. In the following, we reproduce in a straightforward way the results of Refs.~\cite{Clesse:2013jra,Clesse:2010iz,Kodama:2011vs} on the waterfall dynamics and refer the reader to them for the detailed calculation.  

In phase-1, integrating the slow-roll equations gives the field trajectories
\be
\xi^2 = \frac{M^2}{4 \phic \mu_1} \chi.
\ee
If we define the $e$-fold time $\NN$ by imposing that $\NN=0$ at the critical point $\phic$, one gets also
\be
\NN = - \frac{\xi \mu_1 \phic}{\Mpl^2},
\ee
so that the total number of $e$-folds in phase-1 is given by
\be
N_1 = \dfrac{\sqrt{\chi_2} \phic^{1/2}  \mu_1^{1/2} M  }{2  \Mpl^2},
\ee
where 
\be
\chi_2 \equiv  \ln  \left( \frac{\phic^{1/2} M}{ 2  \mu_1^{1/2} \psi_{0}}  \right)
\ee
is the value of $\chi$ at the transition point between phase-1 and phase-2.   We focus only on the regime where the temporal minimum of the potential, defined by the ellipse of local minima in the $\psi$ direction, $ \xi = - \psi^2 / M^2 $,    
is not reached by field trajectories before the end of inflation. 

By integrating the slow-roll equations in phase-2, one gets that trajectories satisfy
\be
\xi^2 = \xi_2^2 + \frac{M^2}{8 \phic \mu_1} \left[ \rr e^{2^{(\chi - \chi_2)}} -1 \right]
\ee
where 
\be
\xi_2 \equiv - \frac{M \sqrt{\chi_2}}{2 \sqrt{\mu_1 \phic} }
\ee
is the value of $\xi$ at the transition between the two phases.   Inflation ends when the slow-roll approximation breaks for the field $\psi$, at
\be
\xiend  = - \frac{M^2}{8 \Mpl^2} .
\ee
Assuming that $|\xiend| \gg |\xi_2|$ (which is valid in the mild waterfall case with $N\gg 50$) the total number of $e$-folds in phase-2  is well approximated by
\be
N_2 \simeq  \frac{M \mu_1^{1/2} \phi^{1/2} }{4 \Mpl^2 \chi_2^{1/2}}.
\ee

\subsubsection{Power Spectrum of Curvature Perturbations}

The power spectrum of curvature perturbations has been calculated numerically by integrating both the classical exact homogeneous dynamics and the linear perturbations.  For this purpose, we have used the method of~\cite{Ringeval:2007am}, similarly to what was done in~\cite{Clesse:2013jra} in the case of a long waterfall phase.  As a cross check, we have used the $\delta N$ formalism~\cite{Lyth:2005fi,Langlois:2008vk,Sugiyama:2012tj} that allows to derive good analytical approximations for the power spectrum of curvature perturbations.   

In the $\delta N$ formalism, the scalar power spectrum at a wavelength mode $k$ is given by
\be
\mathcal P_\zeta (k) = \frac{H_k^ 2}{4 \pi^2 } (N_{,\psi}^ 2 + N_{,\phi}^ 2 ),
\ee
where $N_{,\phi}$ and $N_{, \psi}$ are defined as 
\be
N_{,\phi} \equiv \frac{\partial N_i^ f}{\partial \phi_k^ i}, \hspace{1cm}
N_{,\psi} \equiv \frac{\partial N_i^ f}{\partial \psi_k^ i},
\ee
and $N_i^ f$ corresponds to the number of $e$-folds realized between an initially flat hypersurface at the time $t_k$ defined such that $k /a(t_k) = H(t_k)$, and a final surface of uniform energy density, which we take to be the hypersurface of $\xi = \xiend$\footnote{This choice does not correspond exactly to a hypersurface of constant density $\rho_{\rr{end}}$, but as explained in~\cite{Clesse:2013jra} the difference in $e$-folds between the two hypersurfaces is marginal and can be safely neglected.}.   

One therefore needs to calculate how the number of $e$-folds varies in order to reach $\xi = \xiend$ when the initial field values at the time $t_k$ are slightly shifted.  Actually, one can show that the phase-2 is effectively single field so that one can focus only on modes leaving the Hubble radius during phase-1.   It is useful to decompose $N_{,\phi} = N_{,\phi}^1 + N_{,\phi}^2 $ and $N_{,\psi} = N_{,\psi}^1 + N_{,\psi}^2$, where the first and second terms correspond to the variation of the number of $e$-folds realized in phase-1 and phase-2 respectively.   The number of $e$-folds in phase-1 from arbitrary field values 
$(\xi_{\rr i}, \chi_{\rr i})$ is given by
\be
N_1 = - \frac{\mu_1 \phic (\xi_{2 \rr i} - \xi_{\rr i})}{\Mpl^ 2}
\ee
where $\xi_{2 \rr i} \equiv - \sqrt{\xi_{\rr i}^ 2 + M^ 2(\chi_2 - \chi_{\rr i})/(4 \mu_1 \phic)}$ is the value of $\xi$ at the transition between phase-1 and phase-2.  One recovers $\xi_{2 \rr i} = \xi_2$ when setting $\xi_{\rr i}^2 = M^2 \chi_{\rr i} / (4 \phic \mu_1)$.   It has been shown~\cite{Clesse:2013jra} that the number $e$-folds in phase-2 from field values $(\xi_{2i}, \chi_2)$  gives a subdominant contribution to $N_{,\phi}$ and $N_{,\psi}$.   The importance of the errors induced by neglecting $N_{,\phi}^2$ and $N_{,\psi}^2$ will be nevertheless discussed later.   Under those conditions, one therefore gets
\be
N_{,\phi} \simeq \frac{\mu_1}{\Mpl^ 2}, \hspace{10mm} N_{,\psi} \simeq \frac{M^ 2}{8 \Mpl^ 2 \xi_2 \psi_k},
\ee 
with $ \psi_k = \psi_0 \exp \chi_k$, $\chi_k = 4 \phic \mu_1 \xi_k^2 / M^2$ and $\xi_k = - \Mpl^2 (N_1 + N_2 - N_k) / (\mu_1 \phic )$.  For $10 \lesssim N_k \lesssim 50$, it follows that $N_{,\phi} \ll N_{,\psi}$.  The power spectrum amplitude then can be approximated by
\be \label{eq:Pzeta_exit_in_phase1}
{\cal P}_\zeta (k) \simeq \frac{\Lambda M^2 \mu_1 \phic}{192 \pi^2  \Mpl^6 \chi_2 \psi_k^2}  
\,.
\ee
For the mode exiting the Hubble radius exactly at the transition between phase-2 and phase-1, one can obtain
\be
{\cal P}_\zeta (k, t_k \simeq t_{1,2}) \simeq \frac{\Lambda \mu_1^{2}}{48 \pi^2 p^2 \Mpl^6 \chi_2 }\,
\,,
\ee
whereas for modes exiting the horizon deeper in phase-1, one obtains an exponential increase of the power spectrum amplitude
\begin{eqnarray}
{\cal P}_\zeta (k)  & = & {\cal P}_\zeta (k, t_k \simeq t_{1,2}) \nonumber\\ 
& \times & \exp \left[ 2 \chi_2 \left( 1 - \frac{{(\Nend -N_k)}^2}{N_1^2}   \right)   \right]\,.
\label{P:exponential}
\end{eqnarray}
It is maximal at the critical point of instability.  The mild-waterfall therefore induces a broad peak in the scalar power spectrum for modes leaving the horizon in phase-1 and just before the critical point.   The maximal amplitude for the scalar power spectrum is given by
\be \label{eq:Pzetaapprox}
{\cal P}_\zeta (k_{\phic}) \simeq \frac{\Lambda M^2 \mu_1 \phic}{192 \pi^2  \Mpl^6 \chi_2 \psi_0^2 } \,.
\ee
Depending of the model parameters, the curvature perturbations can exceed the threshold value for leading to the formation of PBH.   

This calculation was performed assuming that $\psic = \psi_0 $.  It is important to remark that for values of $\psi_0$ and $\Lambda$ given by Eqs.~(\ref{eq:psi0}) and~(\ref{eq:Lambda}), one gets that $N_1$, $N_2$ and the amplitude of the scalar power spectrum depend on a concrete combination of the parameters, $\Pi \equiv M (\phic \mu_1)^{1/2} /\Mpl^2$, plus some dependence in $\chi_2$.  But $\chi_2$ itself depends only logarithmically on $\Pi$.  As a result, $\chi_2$ varies by no more than 10\% for relevant values of $\Pi^2$.  The parameters $\phic$, $\mu_1$ and $M$ appear to be degenerate and all the model predictions only depend on the value given to $\Pi$.    Nevertheless, Eq.~(\ref{eq:Pzetaapprox}) implicitly assumes that field values are strongly sub-Planckian.  In the opposite case, when $\phic \sim M \sim \Mpl$, we find important deviations and the numerical results indicate that the waterfall is longer by about two $e$-folds and that the power spectrum is enhanced by typically one order of magnitude, compared to what is expected for sub-Planckian fields and for identical values of  $\Pi^2$.

\begin{figure}[!ht]
\begin{center} 
\includegraphics[height=50mm]{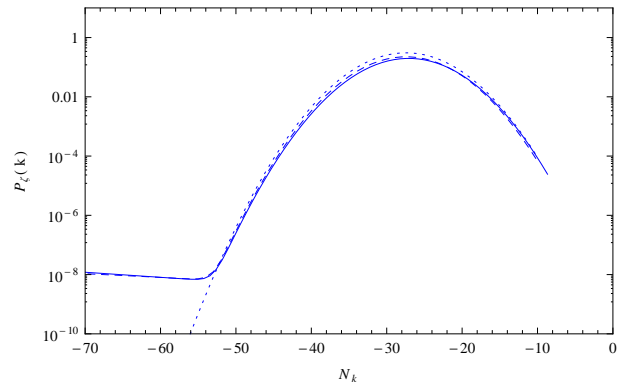}
\caption{\label{fig:Pzetadiffmethods}
Power spectrum of curvature perturbations for parameters $M = \phic = 0.1 \Mpl$, $\mu_1 = 3 \times 10^5 \Mpl$.  The solid curve is obtained by integrating numerically the exact multi-field background and linear perturbation dynamics.  The dashed blue line is obtained by using the $\delta N$ formalism.  The dotted blue line uses the $\delta N$ formalism with the approximation of Eq.~(\ref{eq:Pzeta_exit_in_phase1}). 
}
\end{center}
\end{figure}

As a comparison between numerical and analytical methods, we have plotted in Fig.~\ref{fig:Pzetadiffmethods} the power spectrum of curvature perturbations for $\Pi^2 = 300 $ and sub-Planckian fields, by using the analytical approximation given by Eq.~(\ref{eq:Pzeta_exit_in_phase1}), by using the $\delta N$ formalism including all terms (i.e. the contributions from phase-1 and phase-2) in $N_{,\phi}$ and $N_{,\psi}$, and by integrating numerically the exact dynamics of multi-field perturbations.   As expected we find a good qualitative agreement between the different methods.   Nevertheless, one can observe about $20 \%$ differences when using the analytical approximation, which actually is mostly due to the fact that $N^2_{,\psi}$ has been neglected.   In the rest of the paper, we use the numerical results for a better accuracy.  

\begin{figure}[!ht]
\begin{center} 
\includegraphics[height=50mm]{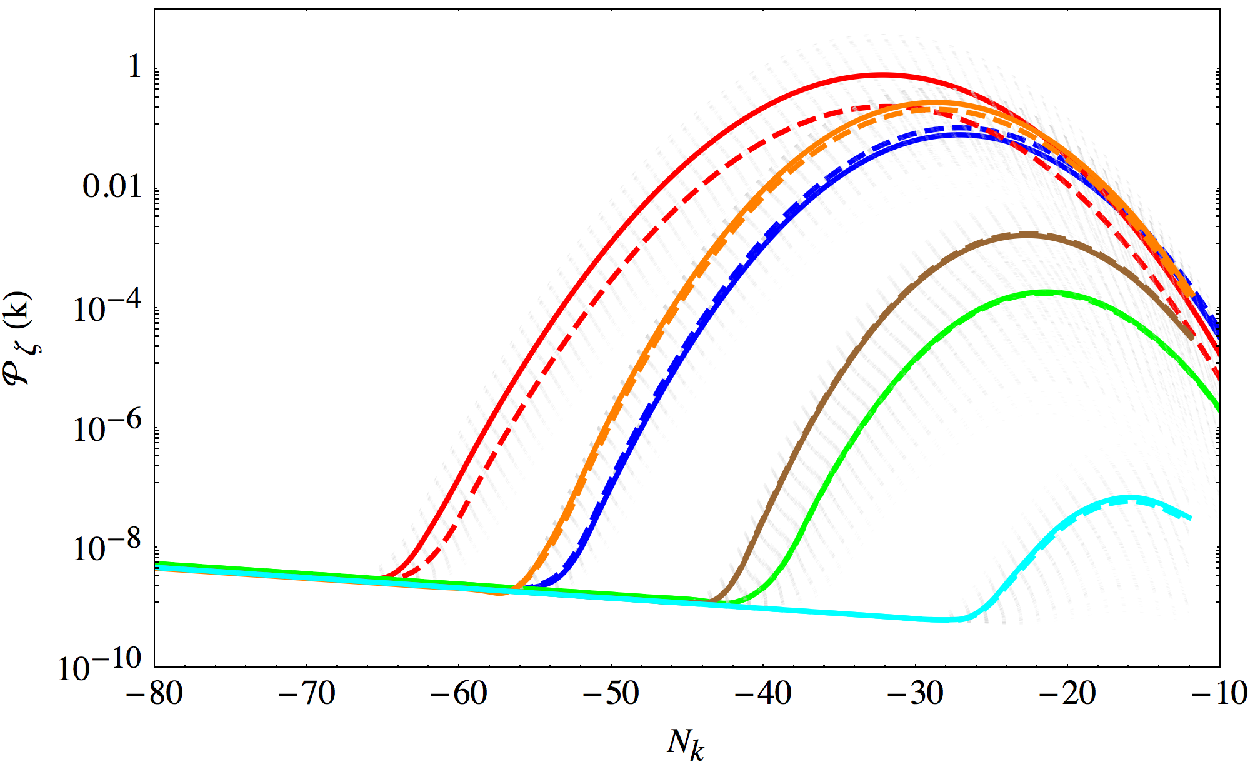}
\caption{\label{fig:Pzeta}
Power spectrum of curvature perturbations for parameters values $M = 0.1 \Mpl$, $\mu_1 = 3 \times 10^5 \Mpl$ and $\phic = 0.125 \Mpl $ (red), $\phic = 0.1 \Mpl $ (blue) and $\phic = 0.075 \Mpl $ (green), $\phic = 0.1 \Mpl $ (blue) and $\phic = 0.05 \Mpl $ (cyan).  Those parameters correspond respectively to $\Pi^2 = 375 / 300 / 225/150$.  The power spectrum is degenerate for lower values of $M,\phi$ and larger values of $\mu_1$, keeping the combination $\Pi^2$ constant.  For larger values of $M, \phic$ the degeneracy is broken: power spectra in orange and brown are obtained respectively for $ M = \phic = \Mpl$ and $ \mu_1 = 300 \Mpl / 225 \Mpl$.  Dashed lines assume $\psic = \psi_0$ whereas solid lines are obtained after averaging over 200 power spectra obtained from initial conditions on $\psic$ distributed according to a Gaussian of width $\psi_0$.  The power spectra corresponding to these realizations are plotted in dashed light gray for illustration.  The $\Lambda$ parameter has been fixed so that the spectrum amplitude on CMB anisotropy scales is in agreement with Planck data.  The parameter $\mu_2 = 10 \Mpl$ so that the scalar spectral index on those scales is given by $\ns = 0.96$.  
}
\end{center}
\end{figure}

In Figure~\ref{fig:Pzeta}  the power spectrum of curvature perturbations has been plotted for different values of the parameters.  This shows the strong enhancement of power not only for the modes exiting the Hubble radius in phase-1, but also for modes becoming super-horizon before field trajectories have crossed the critical point.   One can observe that if the waterfall lasts for about 35 $e$-folds then the modes corresponding to $ 35 \lesssim N_k \lesssim 50 $ are also affected.   As expected one can see also that the combination of parameters $\Pi$ drives the modifications of the power spectrum.   We find that it is hard to modify independently the width, the height and the position of the peak in the scalar power spectrum.  

Finally, for comparison, the power spectra assuming $\psi_c = \psi_0 $ and averaging over a distribution of $\psi_c$ values are displayed.  They nearly coincide for $\Pi^2 \lesssim 300$ but we find significant deviations for larger values.

\section{Formation of Primordial Black Holes}  \label{sec:PBHformation}

In this section, we calculate the mass spectrum of PBH that are formed when $\mathcal O(1)$ curvature perturbations originating from a mild-waterfall phase re-enter inside the horizon and collapse.   Furthermore we show that the abundance of those PBH can coincide with dark matter and can evade the observational constraints mentioned in Sec.~\ref{sec:PBHconstraints}.  

The mass of a PBH whose formation is associated to a density fluctuation with wavenumber $k$, exiting the Hubble radius $| N_k†|$ $e$-folds before the end of inflation, is given by~\cite{GarciaBellido:1996qt} \be \label{eq:Mk}
M_k = \frac{\Mpl^2}{H_k}  \rr e^{-2 N_k} ,
\ee
where $H_k$ is the Hubble rate during inflation at time $t_k$.  In our model, $ H_k \simeq \sqrt{\Lambda / 3 \Mpl^2} $.  

Assuming that the probability distribution of density perturbations are Gaussian, one can evaluate the fraction $\beta$ of the Universe collapsing into primordial black holes of mass $M$ at the time of formation $t_M$ as~\cite{Harada:2013epa}
\ba
\beta^{\rr{form}}(M) & \equiv &  \left. \frac{\rho_{\rr{PBH}} (M) }{\rho_{\rr{tot}}} \right|_{t=t_M} \\ 
 & = & 2 \int_{\zeta_c}^{\infty} \frac{1}{\sqrt{2 \pi} \sigma} \rr e^{- \frac{\zeta^2}{2 \sigma^2}} \dd \zeta \\
& = &  \rr{erfc} \left( \frac{\zeta_c}{\sqrt 2 \sigma}  \right) \label{eq:betaform}.
\ea
In the limit where $\sigma \ll \zeta_c$, one gets
\be \label{eq:betaapprox}
\beta^{\rr{form}}(M) = \frac{\sqrt 2 \sigma}{ \sqrt \pi \zeta_c} \rr e^{- \frac{\zeta_c^2}{2 \sigma^2 }}.
\ee
The variance $\sigma$ of curvature perturbations is related to the power spectrum through $\langle \zeta^2 \rangle = \sigma^2 = \mathcal P_\zeta (k_M) $, where $k_M$ is the wavelength mode re-entering inside the Hubble radius at time $t_M$.   

For the study of PBH formation in our model, the range of masses has been discretized and the value of $\beta^{\rr{form}}$ has been calculated for mass bins $\Delta M$.  This corresponds to PBH formed by the density perturbations reentering inside the horizon between $t_M$ and $t_{M + \Delta M}$.  We have considered mass bins whose width corresponds to one $e$-fold of expansion between these times, i.e.  $\Delta N_k= \Delta \ln k =  (\Delta \ln M )/2 = 1$.  This is sufficiently small for the power spectrum to be considered roughly as constant on each bin.   Thus one gets  $\sigma(M_k) = \sqrt{\mathcal P_\zeta(k) \Delta \ln k}$.  

In the following, we use Eq.~(\ref{eq:betaform}) instead of Eq.~(\ref{eq:betaapprox}), so that the results are accurate when $\sigma \gtrsim \zeta_c$.
Calculating $\zeta_c$ has been the subject of intensive studies~\cite{Nakama:2013ica,Harada:2013epa,Niemeyer:1999ak,Niemeyer:1997mt,Shibata:1999zs,Musco:2004ak,Polnarev:2006aa,Musco:2008hv,Musco:2012au}, using both analytical and numerical methods.  An analytical estimate has been determined recently using a three-zone model to describe the PBH formation process~\cite{Harada:2013epa}.  For a given equation of state parameter $w$ at the time of formation, it is given by
\be
\zeta_c = \frac 1 3 \ln \frac{3 (\chi_a - \sin \chi_a \cos \chi_a)}{2 \sin^3 \chi_a}~,
\ee
with $\chi_a = \pi \sqrt w /(1 + 3 w)$.  In the present scenario, PBH are formed in the radiation dominated era and one can set $w =1/3$, leading to $\zeta_c\simeq 0.086$.   Different values have been obtained with the use of numerical methods, but it seems a general agreement  
that $\zeta_c $ lies in the range $0.01 \lesssim \zeta_c \lesssim 1$ (see Fig.~3 of ~\cite{Harada:2013epa} for a comparison between different methods).  For the sake of generality, $\zeta_c$ will be kept as a free parameter.  

\begin{figure}[!ht]
\begin{center} 
\includegraphics[height=60mm]{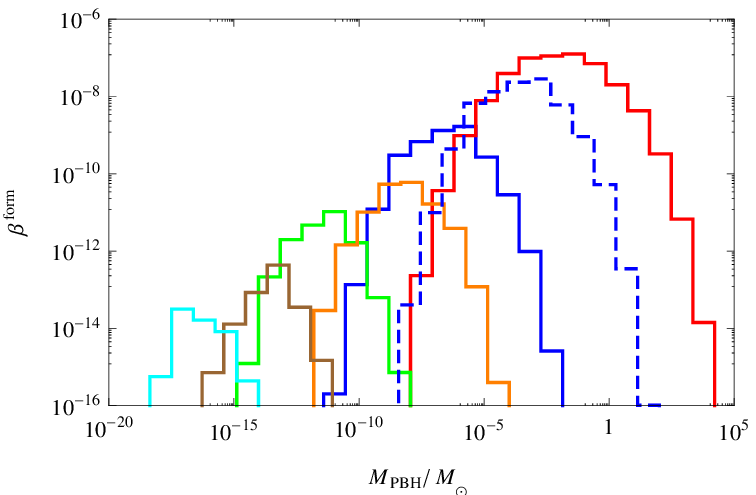}
\includegraphics[height=57mm]{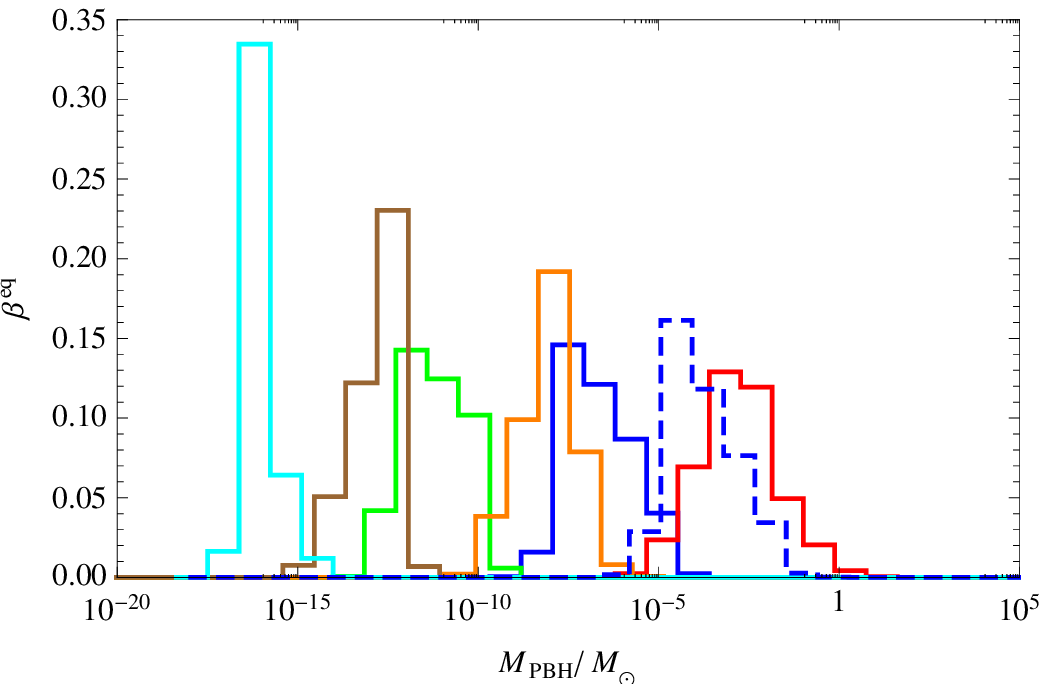}
\caption{\label{fig:betaspectrum} \label{fig:betaspectrumzeq}
PBH abundances at the time of formation $\beta^{\rr{form}}(M)$ (top panel) and at matter-radiation equality $\beta(M, N_{eq})$ (bottom panel).  The color scheme corresponds to the parameters given in Fig.~\ref{fig:Pzeta}.  The blue dashed curve is obtained for $\Pi^2 = 300$ but with $M = \phic = 0.01 \Mpl$ and $\mu_1 = 3 \times 10^8 \Mpl $, illustrating that PBH masses can be made arbitrarily large for a given value of $\Pi^2$.   The critical curvature $\zeta_c$ has been set so that the right amount of dark matter has been produced at matter-radiation equality.   Values of $\zeta_c$ are reported in Table~\ref{tab:omegaPBH}
}
\end{center}
\end{figure}

In our scenario of mild waterfall, the peak in the power spectrum of curvature perturbations is broad and covers several order of magnitudes in  wavenumbers.  Therefore, instead of a   distribution of black holes that would be close to monochromatic, which is easy to evolve in the radiation era, one expects that PBH have a broad mass spectrum and form at different times in the radiation era.   Since the energy density associated to PBH of mass $M$ decreases like $\sim a^{-3}$ due to expansion,  the contribution of PBH to the total energy density in the radiation era grows like $\sim a$.  As a result, at the end of the radiation era, PBH with low masses, forming earlier, contribute more importantly to the total energy density than more massive ones, forming later, given identical values of $\beta^{\rr{form}}$.    

In addition, one must consider that a fraction the PBH is absorbed by the formation of more massive ones at later times. This merging process is subdominant as long at $\beta(M) \ll 1$, but has nevertheless been considered in the 
calculation of the ratio $\Omega_{\rr{PBH}}(z_{eq}) \equiv \rho_{\rr{PBH}}(z_{eq}) / \rho_{\rr{tot}}(z_{eq})$ at matter-radiation equality.  

Taking account those considerations, during the radiation-dominated era, the fraction of the Universe that has collapsed into primordial black-holes of mass $M_k$  evolves as 
\be \label{eq:dbetadN}
\frac{\dd \beta(M_{k},N(t))}{\dd N} =  \beta(M_k,N(t))  \left[1 -  \beta^{\rr{form}}(M_{t} ) \right].
\ee
The first term is due to cosmic expansion 
 and the second term represents the fraction of PBH of mass $M_k$ that are absorbed by the formation of PBH with larger mass $M_t$ at the time $t$.  Since we adopt a discretization in $\ln M$ (and equivalently in $\ln k$ and $N_k$), it is convenient to use the $e$-fold time $N(t)$ instead of the cosmic time.   Note that we neglect evaporation through Hawking radiation since it is relevant only for PBH with very low masses that are formed immediately after inflation.   These are very subdominant in our model due to the duration of the waterfall.   In order to get $\beta^{\rr{eq}} \equiv \beta(M_k,N(t_{\rr{eq}}))$, this equation must be integrated over cosmic history, from the time of PBH formation until matter-radiation equality.   For all the considered curvature power spectra, the formation of PBH stops before $N_{\rr{eq}}$ (corresponding to $\ln (a_{\rr{eq}} / a_0) \simeq - 8 $), since the variance of curvature perturbations can be close or overpass the threshold value only in the range $ -40 \lesssim - N_k \lesssim 10 $.  

The total density of PBH at radiation-matter equality is obtained by integrating $\beta^{\rr{eq}}$ over masses:
\be \label{eq:OmPBH}
\Omega_{\rr{PBH}}(z_{\rr{eq}})  = \int_0^{M_{t_{\rr{eq}}}} \beta(M,N_{\rr{eq}} )  \dd \ln M .
\ee
Eqs.~(\ref{eq:dbetadN}) and~(\ref{eq:OmPBH}) have been solved numerically using bins $\Delta N = 1$, corresponding to $\Delta \ln M = 2 $.  At matter-radiation equality one has 
$\Omega_{\rr{matter}}(t_{eq}) = 0.5 $ and PBH constitute the totality of the dark matter if $\Omega_{\rr{PBH}}(t_{eq}) \simeq 0.42$, the rest coming from baryons.  For simplicity we have neglected the matter contribution to the Universe expansion in the radiation era.  This effect is only important close to matter-radiation equality, when all PBH are formed, and it is expected to be compensated by a small variation of $\zeta_c$.

For the parameter sets considered in Fig.~\ref{fig:Pzeta}, we have found the value of $\zeta_c$ that give rise to the right amount of dark matter.   They are reported in Table~\ref{tab:omegaPBH}.  This must not be seen as an accurate result, because the matter contribution to the Universe's expansion is not accounted for in Eq.~(\ref{eq:dbetadN}) whereas it is not negligible in the last few $e$-folds before reaching matter-radiation equality.   This effect reduces the value of $\beta^{\rr{eq}} $, which must be compensated by a lower value of $\zeta_c $ to get the right amount of dark matter (thus values $\zeta_c /\zeta_{c\rr{fid} } $ of a few tens can still be seen as realistic).   The corresponding $\beta^{\rr{form}}$ and $ \beta^{\rr{eq}}$ functions are represented in Fig.~\ref{fig:betaspectrum}.  As expected, $\beta^{\rr{form}}$ is much smaller than $ \beta^{\rr{eq}}$ (remember that $\beta \propto a$).  It takes larger values for massive black holes that are produced after a long waterfall phase, since they are formed later and thus if we identify them to dark matter, $\beta$ must grow up to values of $\mathcal O(0.1) $ in a fewer number of $e$-folds.  The amplitude of the peak in $\beta^{\rr{form}}$ is controlled by $\Pi^2$, but also by the value of $\mu_1$.  Indeed, to larger values of $\mu_1$ correspond lower energy scales for inflation, and it results that PBH are formed at higher redshifs.    
\begin{table}[ht!]
\begin{center} \label{tab:omegaPBH}
\begin{tabular}{|c|c|c|c|c|c|}
\hline
 $\Pi^2 (\mu_1, M, \phic \, \rr{in} \, \Mpl) $  & $\zeta_c / \zeta_{c \rr{fid}} $  & $\mu^{dist}$ &$y^{dist}$ \\
\hline
 $375 \, (3 \times 10^5, 0.1, 0.125) $ & $102.4 $  & $2.7 \times 10^{-3}$  & $1.3 \times 10^{-5}$ \\
 $300 \, (3 \times 10^5, 0.1, 0.1) $ & $22.05 $ & $1.6 \times 10^{-7}$  & $5.2 \times 10^{-9}$\\
 $300 \, (3 \times 10^8, 0.01, 0.01) $ & $20.2 $ & $1.6 \times 10^{-7}$  & $5.2 \times 10^{-9}$\\
 $300 \, (300, 1, 1) $ & $ 49.60$  & $2.6 \times 10^{-6}$ & $5.6 \times 10^{-9}$\\
 $225 \, (3 \times 10^5, 0.1, 0.075)  $ & $2.337$  & $4.9 \times 10^{-9}$ & $5.4 \times 10^{-9}$ \\
 $225 (225, 1, 1) $ & $6.060$  & $4.0 \times 10^{-9}$ & $4.6 \times 10^{-9}$ \\
 $150 \, (3 \times 10^5, 0.1, 0.05) $  & $0.0567$  & $4.0 \times 10^{-9}$ & $4.6 \times 10^{-9}$\\
\hline 
\end{tabular}
\end{center}
\caption{ Critical value $\zeta_c$ of curvature fluctuation (2nd column) leading to PBH formation with $\Omega_{\rr{PBH}} (z_{\rr{eq}}) = 0.42$ at matter radiation equality, for several sets of the model parameters (1st column).  The fiducial value is $\zeta_{c \rr{fid}} = 0.086$, according to the three-zone model of~\cite{Harada:2013epa}.  In 3rd and 4th columns are reported the corresponding distortion  $\mu$ and $y$ parameters (see Sec.~\ref{sec:distortions}). }  
\end{table}

The mass range for PBH is very broad, $10^{-20} \msun \lesssim \mPBH \lesssim 10^{5} \msun$.  But given one set of parameter, the mass spectrum typically covers 3-5 order of magnitudes at matter-radiation equality.  Given $\Pi^2$, we find that PBH can be made arbitrarily massive by increasing $\mu_1$ and reducing $M$ and $\phic$.  This lowers the energy scale of inflation and thus increases PBH masses, but this does not affect importantly the shape of the mass spectrum.  Therefore it is easy to find parameters for which the mass spectrum peaks in the range where there is no solid observational constraints.  It is also possible that the peak in the mass spectrum is located on planet-like masses at recombination (so that CMB distortion constraints are satisfied), but evade micro-lensing limits of PBH abundances if merging induces their growth by more than two or three orders of magnitudes during cosmic history.    

Finally, the width of the peak in  $ \beta^{\rr{eq}}$ is reduced for lower values of $\Pi^2$, as expected given that it is related to the broadness of the peak in the scalar power spectrum.  It is therefore possible, in principle, to control this width, but note that the range where $\Pi^2$ can vary is rather limited by the value of  $\zeta_c $, which needs to be realistic.   

\section{Constraints on the inflationary potential}  \label{sec:infconstraints}

Given the strong modifications of the power spectrum of curvature perturbations when varying the potential parameters, modifications that are amplified when considering the mass spectrum of PBH, the scenario we propose predicts that the inflation potential parameters obey to very specific combinations.  As already mentioned, $\mu_1$, $\phic$ and $M^2$  are degenerate and we thus can consider their combination within a unique parameter $\Pi$.  Nevertheless, this is only valid for sub-Planckian fields, and therefore we have also calculated scalar power spectra and BPH mass spectra for the specific values $M = \Mpl, \phic = 0.1 \Mpl$ and $\phi = M = \Mpl$.

First, we find that values
\be
\Pi^2 \equiv \frac{M^2 \phic \mu_1}{\Mpl^4} \gtrsim 400 
\ee
are not allowed observationnally, because the waterfall phase is too long and the increase of power arises already on CMB anisotropy scales.  On the other hand, for values 
\be
\Pi^2 \lesssim 200
\ee
the curvature perturbations are not sufficiently enhanced during the waterfall and the model cannot produce the right abundances of PBH, except if $\zeta_c$ takes unrealistically low values.  Then, for $\Pi^2 \gtrsim 350$ we find that $\zeta_c $ must be much larger than unity, which is also unrealistic.  It results that the model works for very specific values of $\Pi^2$, within the range $200 \lesssim \Pi^2 \lesssim 350$. 

Since $\Lambda \propto \Mpl^6/\mu_1^2$, this bound on $\Pi^2$ gives a maximal energy scale for inflation if the field values are restricted to be sub-planckian, or close to the Planck mass (which is expected in SUSY hybrid models to avoid important corrections of supergravity),
\be
\Lambda^{1/4} \lesssim 8 \times 10^{-3} \Mpl \simeq  2 \times 10^{16} \,\rr{GeV}.
\ee
This scale is close to the GUT energy scale.  From the scalar power spectrum only, there is no lower bound on the energy scale of inflation (other than the usual BBN constraints).  It can be arbitrarily low if the parameters $M$ and $\phic$ are much lower than the planck mass.  The maximal value for the tensor to scalar ratio (requiring Planck-like values of $M$) is given by
\be
r = 16 \epsilon_1 \simeq \frac{8 \Mpl^2}{\mu_1^2} \lesssim 0.08,
\ee
it is somewhat below the present limits, but in the range of detectability of future CMB polarization experiments like COrE+\cite{Bouchet:2011ck}.

A similar bound arises from the observational constraints on PBH abundances.  Values of $\Pi^2 \gtrsim 400 $ combined with sub-Planckian fields will generate PBH with masses larger than $\msun$ at matter-radiation equality, which is ruled out by the limits on CMB distortions.  Furthermore, as shown in Sec.~\ref{sec:distortions}, additional distortions are produced in this case because of higher power on Silk-damped scales and they should have been detected by FIRAS.

\section{Embedding in more realistic inflation models}  \label{sec:embedding}

The effective potential given in Eq.~(\ref{eq:potential}) has a tiny slope and a negative curvature in the valley direction, close to the critical point of instability.  This is opposite to the original hybrid model where the curvature is positive.  This potential can nevertheless arise in the framework of hybrid models having a flat direction lifted up by logarithmic radiative corrections.  The most famous examples are the supersymmetric F-term~\cite{Dvali:1994ms} and D-term models~\cite{Binetruy:1996xj,Halyo:1996pp}.  After a Taylor expansion around $\phic$, the Coleman-Weinberg logarithmic corrections indeed lead to a positive linear term and a negative quadratic term in the effective potential.  In this section, we examine whether our model can be embedded in realistic high-energy framework, considering the parameter space producing a mild waterfall phase lasting a few tens of $e$-folds.

\subsection{F-term supersymmetric model}
The F-term hybrid model is based on the superpotential 
\be
W^{\rr{F-term}} = \kappa S \left( \bar \Phi \Phi - M^2 \right) ,
\ee
where $S$ is a singlet chiral superfield and $\bar \Phi, \Phi$ is a pair of chiral superfields charged under $U(1)_{B-L} $.  Along the D-flat direction, the SUSY vacuum is reached at 
$\langle S \rangle = 0 $ and $  | \langle \Phi \rangle  | =  | \langle \bar \Phi \rangle  | = M  $.  One can introduce the real canonically normalized inflaton field $\phi \equiv \sqrt 2 |S|$.  The effective potential along the flat direction $\bar \Phi = \Phi = 0$ is then given by
\begin{eqnarray}
V^{\rr{Fterm}} (\phi) & = &  \kappa^2 M^4 \left\{1 + \frac{\kappa^2 \bar N}{32 \pi^2} \left[ 2 \ln \frac{\kappa^2 M^2 x}{Q^2} \right. \right. \nonumber \\ 
& + & \left. \left.  (x+1)^2 \ln \left(1 + \frac 1 x \right) + (x-1)^2 \ln \left(1 - \frac 1 x \right)  \right]  \right. \nonumber \\
& - &  \left.   \frac{a_S \phi}{\sqrt 2 \kappa M^2}  \right\}.
\end{eqnarray}
The second term represents the radiative corrections to the tree-level potential, $x\equiv \phi^2 / 2 M^2$, $Q$ is a renormalization scale and $\bar N $ is the dimensionality of the representation to which $\bar \Phi, \Phi$ belong.  The third term is a possible soft SUSY-breaking term, which we first consider to be negligible.   We have neglected supergravity corrections to the potential.  They are negligible as long as the fields are much lower than the reduced Planck mass but could lift up the potential for Planck-like values.   The instability point is located at $\phi_c = \sqrt 2 M $.   

After expanding in Taylor series the F-term potential close to $\phic$, one recovers the potential of Eq.~(\ref{eq:potential}) and one can identify (in the limit of vanishing soft-term)
\be
\Lambda = \kappa^2 M^4,
\ee
\be
\frac{1}{\mu_1} = \frac{\kappa^2 \bar N \ln 2 }{8 \pi^2},
\ee
\be
\frac{1}{\mu_2^2} = \frac{\kappa^2 \bar N}{32 \pi^2} \left( \frac{3}{2} - \ln 2 \right).
\ee
Two of the three F-term parameters can then be determined by requiring that the scalar power spectrum amplitude and spectral index are in agreement with CMB observations.  The third parameter can be determined with the requirement of a mild waterfall, which is translated in constraints on $\Pi^2$.    However, we find that in order to have $\Pi \sim 200$ as well as $M \lesssim \Mpl$ to avoid supergravity corrections, then one must satisfy $\bar N \gtrsim \mathcal O (10^5)$ which is very unrealistic since this parameter denotes the dimensionality of the representation of the superfields $\bar \Phi, \Phi$.   

Therefore it is not possible to embed our effective potential in the standard version of the F-term scenario.  Nevertheless, if the potential is protected by a gauge symmetry, one can allow the fields to take values of the order of the reduced Planck mass.  Then one can satisfy $\Pi^2 \sim 200 $ and $\bar N \sim \mathcal O (1)$ simultaneously.  

Another possibility is that the soft SUSY-breaking term plays the role of reducing the slope of the potential close to the instability point, while the curvature is unchanged.  This extends the duration of the waterfall phase, and one can identify
\be
\frac{1}{\mu_1} = \frac{\kappa^2 \bar N \ln 2 }{8 \pi^2} - \frac{a_S}{\sqrt 2 \kappa M^2}
\ee
 However, if one imposes that $M$ to be lower than the Planck mass and $\bar N \sim \mathcal O(1)$, then the soft term in $\mu_1$ needs to be fine tuned to the term arising from radiative corrections, at least at the percent level.  
  
Finally, let us mention the possibility of a next-to-minimal form for the K\"ahler potential 
 \be
 K =  | S |^2 + k_S \frac{|S|^4}{4 \mpl^2},
 \ee
 where the parameter $k_S$ can be either positive or negative.  In the latter case, this induces a negative mass term $- 3 k_S H^2 |S|^2$ in the effective potential~\cite{Garbrecht:2006az}, and one could identify $ \mu_2^2 = \Mpl^2 / k_S $, the other contributions to $\mu_2$ being subdominant.  This new parameter controls the value of the spectral index.  As a result, larger values of $\mu_1$ and sub-Planckian field values are allowed for the F-term model.   For $\Pi^2 \simeq 200$, this gives
 \be
 \kappa \simeq 5 \times 10^{-17}, \hspace{10mm} M \simeq 1.6 \times 10^{-11} \Mpl,
 \ee  
 but then the energy scale of inflation is very low, at the GeV scale, and the PBH are 
 much too massive, $ M_{\rr{PBH}}\sim 10^{22} M_\odot$.

 \subsection{Loop/D-term inflation}
 
Radiative corrections of a flat direction can give rise to a potential of the form
 \be
 F^{\rr{loop}} = \Lambda \left[ 1+ \lambda_1 \ln \left( \frac{ \lambda_2 \phi^p } {\Mpl^p} \right)   \right].
 \ee
D-term inflation, for instance, belongs to this class.  After expanding around $\phi_c$, one can identify 
\be
\frac{1}{\mu_1} = \frac{p \lambda_1 \Mpl }{\phic}, \frac{1}{\mu_2^2} = - \frac{p \lambda_1 \Mpl^2 }{2 \phic^2} 
\ee
that does not depend on $\lambda_2$.  The effect of $p$ can also be incorporated by a redefinition of $\lambda_1 \rightarrow p \lambda_1$.  Then the relevant value of $\lambda_1$, $\phi_c $ and $\Lambda$ can be derived from the measurements of the scalar spectrum amplitude and tilt and 
by imposing $ \Pi^2 \sim 200$ .   In particular, one finds that
\be
\Pi^2 = \frac{2 M^2}{(\ns - 1) }   
\ee
and therefore if $\ns = 0.96$ one gets $\Pi^2 = 200$ for $ M \simeq 2 \Mpl$.   We therefore conclude that, as for the F-term model, one requires field values close to the Planck scale.  However, note that only the value of $M$ needs to be at the Planck scale, the critical value $\phic$ can take much lower values, compensated by larger values of $\mu_1$.  Supergravity corrections along the valley can therefore be subdominant.  In the case of the D-term model, the superpotential reads 
\be
W^{\rr{D-term}} = \kappa S \bar \Phi \Phi 
\ee
 and $M^2$ is identified to the Fayet-Iliopoulos term $\xi_{FI} $ in the D-term 
\be
D = \frac g 2 \left( \Phi^2 - \bar \Phi^2  + \xi_{FI} \right).
\ee
One can also identify $\lambda_1 = g^2 / (16 \pi^2) $, $\phic = \sqrt{\xi_{FI}} g/(2 \kappa) $ and  $\Lambda = g^2  \xi_{FI} /2$.  For the scalar power spectrum amplitude to agree with CMB data, in addition to a Planck-like Fayet-Iliopoulos term, one therefore needs a large coupling $\kappa \sim 10^5$ contrary to the usual regime where $\kappa \sim 10^{-2}$.  The coupling $g$ must be sufficiently small to keep $\phic$ sub-Planckian.

 \subsection{Some alternatives}

One can mention some alternatives for the embedding of the effective potential in a realistic scenario:  dissipative effects can be invoked to reduce the scalar spectral index, as proposed in~\cite{BasteroGil:2004tg}, and to allow larger values of $\mu_2$ and $\mu_1$ in loop inflation.  Inflexion point models where inflation terminates with a waterfall phase could be good candidates if the position of the critical point is tuned to be very close to a flat inflexion point.  Another interesting possibility is the natural inflation potential $V(\phi) = \Lambda [1+a  \cos(\phi / f) ]$, coupled to an auxiliary sector that triggers a waterfall phase close to the maximum of the potential, where the potential is flat with a negative curvature.  Such a model has been studied in~\cite{Vazquez:2014uca} assuming instantaneous waterfall.

\section{The seed of supermassive black holes}  \label{sec:seeds}

The center of galaxies is believed to host supermassive black holes (SMBH) of mass going from $10^6 M_\odot$ to $10^9 M_\odot$.   These are thought to descent from less massive BH seeds in quasars at high redshifts.  But the existence of SMBH at redshift up to $z\gtrsim 8$~\cite{Fan:2003wd,Willott:2003xf,2010Natur.467..940L,Iye:2006mb,Oesch:2013pt,2014ApJ...786..108O} remains a mystery.   Their existence as fully formed galaxies before 500 Myr is a challenge for standard $\Lambda$CDM model. It is extremely difficult to form such a massive BH so quickly from stellar evolution, and several proposals suggest that they are built up from smaller BH that act as seeds of the galactic SMBH~\cite{Rubin:2001yw,Bean:2002kx,Duechting:2004dk,2012PhLB..711....1K,2012NatCo...3E1304G,Belotsky:2014kca}.  

Assuming uninterrupted accretion at the Eddington limit, BH seeds of at least $10^3 M_\odot$ are needed at $z \approx 15$.  Our scenario provides a mechanism for the formation of those seeds, in the tail of the BH mass distribution.  As already mentioned, abundance of stellar-mass PBH prior recombination is severely constrained.  However, sub-stellar PBH can grow by merging and it is possible that a significant fraction of the smaller mass BH grow to become intermediate mass black holes (IMBH) at redshift $z\gtrsim 15$.  We expect the mechanisms of accretion for an initial broad spectrum of PBH masses to be very complex, so here we have adopted for simplicity the naive prescription that PBH grow by a factor $f_{\rr{merg}} $ between matter-radiation equality and the late Universe, independently of their mass.  Typically we find that  $f_{\rr{merg}} \gtrsim 10^3 $ for the model to pass both CMB distortions and micro-lensing constraints.  If the PBH spectrum peaks on stellar masses at late time, as for the scenario displayed on Fig.~\ref{fig:PBHconstraints}, we find that $\beta(10^{4}M_\odot) \sim 10^{-5}$.  These rare seeds then can merge and accrete matter to form SMBH.   It is also possible to form SMBH at high redshifts with even more massive PBH seeds in the tail of the spectrum.  So in this scenario, it is easy to get a number of SMBH roughly 1 for $10^{12}$ stellar-mass BH in galaxies, which is expected for a PBH dark matter component.  Our model therefore predicts that i) supermassive black holes should be observed at the center of galaxies at very early times, and ii) their mass distribution should follow a gaussian decrease.    

Moreover, within a generic broad mass distribution of PBH, as produced in our scenario, it is natural that PBH formed in the early universe and cluster during the radiation era~\cite{Chisholm:2005vm,2011PhRvD..84l4031C}. Furthermore it is possible that a significant fraction of the smaller mass BH grow to become intermediate mass black holes (IMBH), which could be responsible for the observed ultra-luminous X-ray sources~\cite{Dewangan:2005mb,Madhusudhan:2005zj,Liu:2013jwd,Bachetti:2014qsa}.

\section{CMB distortions} \label{sec:distortions}

By increasing the amplitude of the scalar power spectrum on scales in the range  $8 \Mpc^{-1} \lesssim k \lesssim 10^4 \Mpc^{-1}$, our model induce potentially observable spectral distortions of the CMB black-body spectrum.  This range of scales corresponds to $-55 \lesssim -N_k \lesssim -48 $ (assuming $N_* \simeq 50$).  For some of the considered parameters, the curvature perturbations are enhanced within that range (see Fig.~\ref{fig:Pzeta}).
 
CMB distortions are produced when the thermal equilibrium is broken due to some energy injection before the recombination, even if electrons and ions can remain in thermal equilibrium due to Coulomb collisions.  The energy injection can be due to several processes, such as the decay or the annihilation of relic particles~\cite{Chluba:2013wsa}.  The Silk damping leads also to some energy injection from the dissipating acoustic waves~\cite{Khatri:2013xwa,Chluba:2013dna,Khatri:2012rt,Chluba:2012we,Chluba:2012gq,Khatri:2013dha}, and the magnitude of this effects is related to the amplitude of the density perturbations.  The increase of power on small scales induced by a mild waterfall can therefore result in higher energy injection to the CMB monopole, resulting is enhanced spectral distortions.   Distortion spectra for a scenario of mild waterfall were calculated in~\cite{Clesse:2014pna} and could be seen by the Primordial Inflation Explorer (PIXIE)~\cite{Kogut:2011xw} and the Polarized Radiation Imaging and Spectroscopy Mission (PRISM)~\cite{Andre:2013afa,Andre:2013nfa}.  Those experiments are expected to improve by several order of magnitudes the present limits on the signal intensity in each frequency bin they probe, with 
 \be
\label{sensitivity:PIXIE}
\delta I_\nu ^{\rr{PIXIE}} = 5 \times 10^{-26} \  \rr{W m^{-2} Sr^{-1} Hz^{-1}}~,
\ee 
and 
\be
\label{sensitivity:PRISM}
\delta I_\nu ^{\rr{PRISM}} = 6.5 \times 10^{-27} \ \rr{W m^{-2} Sr^{-1} Hz^{-1}}.
\ee 
In this section, we have reproduced the results of~\cite{Clesse:2014pna} for our effective potential and for relevant parameters in the context of PBH production. 
For this purpose, we have used a modified version of the \texttt{idistort} template~\cite{Khatri:2013dha}, which solves the Kompannets equations and calculates the spectrum of CMB-distortions.   The code has been modified to allow any shape of the primordial scalar power spectrum.   

Distortions can be of $\mu$-type, $y$-type and intermediate $i$-type depending on when during the cosmic history the thermal equilibrium is broken.  The importance of different types is usually encoded in the so-called $\mu$ and $y$ parameters.  The present limits from COBE-FIRAS are $\mu < 9 \times 10^{-5}  $ and $y < 1.5 \times 10^{-5} $ and the objective of PIXIE/PRISM is to improve this limit by about three orders of magnitudes.  

\begin{figure}
\begin{center} 
\includegraphics[height=55mm]{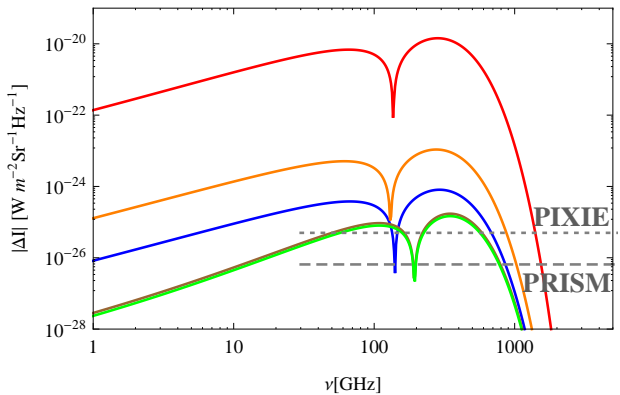}
\caption{\label{fig:distortions}
Total spectrum of CMB distortions for same parameters and colors as in Figs.~\ref{fig:Pzeta} and \ref{fig:betaspectrumzeq}  (brown and green curves are superimposed, undistinguishable from standard inflation with $\ns = 0.961$ and no running).  The  $1\sigma$ limits for PIXIE and PRISM, see  Eqs.~(\ref{sensitivity:PIXIE})~and~(\ref{sensitivity:PRISM}), are also represented.
}
\end{center}
\end{figure}

In addition to distortion spectra, $\mu$ and $y$ values have been calculated for the parameters in Table~\ref{tab:omegaPBH}.  The corresponding spectra are displayed in Fig.~\ref{fig:distortions}.   We find that the distortion signal can be enhanced by several order of magnitudes compared to the standard case where the nearly scale-invariant scalar power spectrum can be extended down to small scales.   As expected the effect is maximal for $\Pi \simeq 300$ whereas the spectrum cannot be distinguished from the standard case when $\Pi \lesssim 200$.   Nevertheless, for $\Pi^2 \sim 220$ the enhancement is about 10\%.   We therefore conclude that if PBH are identified to dark matter and if $\zeta_c$ takes reasonable values, corresponding to $\Pi^2 \sim 200$, the induced spectral distortions pass the present constraints but are sufficiently important to be detected by a PRISM-like experiment.  Our model therefore has a very specific prediction and could be tested with future observations.

\section{Conclusion}  \label{sec:conclusion}

We have proposed a model where dark matter is composed of massive primordial black holes formed in the early Universe due to the collapse of large curvature fluctuations generated during a mild-waterfall phase of hybrid inflation. This regime is transitory between the usual fast-waterfall assumption and the mild-waterfall case with more than 50 $e$-folds of expansion realized after the crossing of the critical instability point of the potential.  In our scenario, the waterfall lasts between 20 and 40 $e$-folds.   The classical field trajectories and the power spectrum of curvature perturbations have been calculated both by using analytical approximations and by solving numerically the exact background and linear perturbation dynamics.  The quantum diffusion close to the instability point has been accounted for by considering and averaging over many possible realizations of the auxiliary field at the instability point, distributed accordingly to the quantum stochastic treatment of this field, whereas the inflaton itself remains classical.  
 
Once the potential parameters are chosen to fit with CMB anisotropy observations, we have shown that a quantity combining the position of the critical instability point, the position of the global minima of the potential and the slope of the potential at the critical point, controls the duration of the waterfall, the peak amplitude and its position in the power spectrum of curvature perturbations.   This parameter therefore controls also the shape of the PBH mass spectrum.  An additional parameter comes from the threshold curvature fluctuation from which gravitational collapse leads to PBH formation when it reenters inside the Hubble radius during the radiation phase.  For realistic values, we have identified the potential parameter ranges leading to the right amount of PBH dark matter at matter-radiation equality.   If PBH masses then grow by merging or accretion, by at least a factor $10^3$, we find that the model can be in agreement with the current constraints on PBH abundances.   In particular, we have identified a scenario where the PBH spectrum peak is centered on sub-solar masses, thus evading CMB distortion constraints, and is then shifted up to stellar-like masses today, thus evading constraints from micro-lensing observations.  This scenario explains the excess of BH candidates in the central region of the Andromeda galaxy.   

Our effective hybrid potential can be embedded in a hybrid model where the slope of the potential in the valley direction is due to logarithmic radiative corrections.  In particular, it was 
found that the above scenario works well for D-term inflation with Planck-like values of Fayet-Iliopoulos term.

Finally we discussed whether PBH in the tail of the distribution can serve as the seeds of the supermassive black holes observed at the center of galaxies and in high-redshift quasars.  Seeds having a mass larger than $10^4 M_\odot$ at redshift $z\sim 15$ are produced and can then accrete matter and merge until they form supermassive black holes (SMBH).  This does not require any specific additional tuning of parameters and is obtained for free from our model, whereas the formation of SMBH at high redshifts is challenging is standard $\Lambda$CDM cosmology.  PBH with intermediate masses are also produced and could explain ultra-luminous X-ray sources.

It is worth mentioning that our scenario leads to specific predictions that could help to distinguish it from other dark matter scenarios in the near future.  First, significant CMB distortions are expected due to the increase of power in the Silk-damped tail of the scalar power spectrum.  We found that they could be detected by PIXIE or PRISM.  Then, a large number of stellar-mass BH candidates in nearby galaxies should be detected by X-rays observations.   Moreover, PBH binaries should emit gravitational waves that could be detected by future gravitational waves experiments such as LIGO, DECIGO and LISA~\cite{2009arXiv0909.1738H,Dolgov:2013pha}.  Finally, X-rays photons that are emitted by PBH at high redshifts should affect the reionization of the nearby intergalactic medium and leave imprints on the 21cm signal from the reionization and the late dark ages~\cite{Tashiro:2012qe}.  In the next decade, 21cm observations by the Square Kilometre Array will constrain the abundance of PBH in the range from $10^2 M_\odot$ to $10^8 M_\odot$, down to a level $ \beta \sim 10^{-9}$, which is enough to rule out our model.  

Further work is certainly needed to understand the role and the importance of mergers between the time of formation and the late Universe.  The fact that the PBH distribution is broad makes this process even more complex than in simpler scenarios.  The clustering of PBH holes, and especially its relation with the quantum diffusion during the inflationary phase, is an open question.  Qualitatively, the scalar power spectrum at the end of inflation varies in different patches of the Universe, having emerged from different realizations of the quantum  diffusion of the auxiliary field at the instability point.  In those patches, the PBH production rate will be different, and therefore will lead to different dark matter abundances.  One thus expects the Universe to be formed of a mixture of large voids and high dark matter density regions, on scales going from a few Mpc up to thousands of Mpc, depending on the inflation potential parameters.  This picture is in some way similar to the Swiss-cheese model, where inhomogeneities can lead to an apparent cosmic acceleration mimicking dark energy.  Holes larger than 35 Mpc are expected to leave distinguishable signatures on the CMB, but smaller sizes are still viable~\cite{Valkenburg:2009iw}.  It will be very interesting but challenging to investigate whether apparent local cosmic acceleration can be obtained from the inhomogeneous structures in our model.

\section{Acknowledgments}

We warmly thank A. Linde, J. Silk and F. Capela for useful comments and discussion.  The work of S.C. is supported by the \textit{Return Grant} program of the Belgian Science Policy (BELSPO).   JGB acknowledges financial support from the Spanish MINECO under grant FPA2012-39684-C03-02 and Consolider-Ingenio ``Physics of the Accelerating Universe (PAU)" (CSD2007-00060).
We also acknowledge the support from the Spanish MINECO's ``Centro de Excelencia Severo Ochoa" Programme under Grant No. SEV-2012-0249.

\bibliography{biblio}

\end{document}